\documentclass[9pt,preprint]{sigplanconf}

\usepackage{amsmath}
\usepackage{color}
\usepackage{graphicx}
\usepackage{colortbl}
\usepackage{booktabs}
\usepackage{float}
\usepackage{xspace}
\usepackage{multirow}
\usepackage{capt-of}
\usepackage{listings}
\usepackage[color=yellow!60,textsize=footnotesize,obeyDraft]{todonotes}
\usepackage{amsmath}
\usepackage{url}
\usepackage{makecell}
\usepackage{ textcomp }

\unless\ifdefined\ARXIV
    \graphicspath{{./media/}}
\fi

\newcommand{\mathspace}[1]{\ensuremath{#1}\xspace}

\newcommand{\instit}[1]{\mathspace{\mathsf{#1}}}
\newcommand{\CR}[1]{\mathspace{C_{\operatorname{#1}}}}
\newcommand{\tuple}[2]{\mathspace{\langle #1, #2 \rangle}}

\newcommand{\aBlock}{\mathspace{\mathit{BB}}}
\newcommand{\aVar}{\mathspace{x}}
\newcommand{\Gen}{\mathspace{\mathit{gen}}}
\newcommand{\Kill}{\mathspace{\mathit{kill}}}

\newcommand{\inst}{\mathspace{\mathit{inst}}}
\newcommand{\seval}{\mathspace{\mathit{seval}}}
\newcommand{\args}{\mathspace{\mathit{args}}}
\newcommand{\flownext}{\mathspace{\mathit{next}}}
\newcommand{\Ref}{\mathspace{\mathit{ref}}}

\newcommand{\op}{\instit{op}}
\newcommand{\memload}{\instit{memload}}
\newcommand{\memstore}{\instit{memstore}}
\newcommand{\call}{\instit{call}}
\newcommand{\ret}{\instit{ret}}
\newcommand{\br}{\instit{br}}

\newcommand{\figlabel}[1]{\label{fig:#1}}

\newcommand{\figref}[1]{Figure~\ref{fig:#1}}

\newcommand{\seclabel}[1]{\label{sec:#1}}

\newcommand{\secref}[1]{Section~\ref{sec:#1}}

\lstnewenvironment{llvmcode}[1][]{\nocaptionrule\lstset{language=LLVM,#1}}{\vspace{-1.1em}}
\newcommand\llvminline[1]{\lstinline[language=LLVM]{#1}}

\long\def \@makecaption #1#2{%
  \addvspace{4pt}
  \if \@caprule
    \hrule width \hsize height .33pt
    \vspace{4pt}
  \fi
  \setbox \@tempboxa = \hbox{\@setfigurenumber{#1.}\nut #2}%
  \if \@dimgtrp{\wd\@tempboxa}{\hsize}%
    \noindent \@setfigurenumber{#1.}\nut #2\par
  \else
    \centerline{\box\@tempboxa}%
  \fi}

\title{Static energy consumption analysis of LLVM IR programs}
\begin{document}

\setlength{\pdfpageheight}{\paperheight}
\setlength{\pdfpagewidth}{\paperwidth}

\conferenceinfo{CONF 'yy}{Month d--d, 20yy, City, ST, Country}
\copyrightyear{20yy}
\copyrightdata{978-1-nnnn-nnnn-n/yy/mm}
\doi{nnnnnnn.nnnnnnn}




\titlebanner{}        
\preprintfooter{}   

\authorinfo{Neville Grech, Kyriakos Georgiou, James Pallister, Steve Kerrison,
Jeremy Morse and Kerstin Eder}{University of Bristol, Merchant Venturers
Building, Woodland Road\\Bristol, BS8 1UB, United
Kingdom}{\{n.grech, kyriakos.georgiou, james.pallister, steve.kerrison, jeremy.morse, kerstin.eder\}@bristol.ac.uk}

\maketitle

\begin{abstract}
Energy models can be constructed by characterizing the energy consumed by
executing each instruction in a processor's instruction set. This can be used to
determine how much energy is required to execute a sequence of assembly
instructions, without the need to instrument or measure hardware.

However, statically analyzing low-level program structures is hard, and the gap
between the high-level program structure and the low-level energy models needs
to be bridged. We have developed techniques for performing a static analysis on the
intermediate compiler representations of a program. Specifically, we target LLVM
IR, a representation used by modern compilers, including Clang. Using these
techniques we can automatically infer an estimate of the energy
consumed when running a function under different platforms, using different
compilers.

One of the challenges in doing so is that of determining an energy cost of
executing LLVM IR program segments, for which we have developed two different
approaches. When this information is used in conjunction with our analysis, we
are able to infer energy formulae that characterize the energy consumption for a
particular program. This approach can be applied to any languages targeting the
LLVM toolchain, including C and XC or architectures such as ARM Cortex-M or XMOS
xCORE, with a focus towards embedded platforms. Our techniques are validated on
these platforms by comparing the static analysis results to the physical
measurements taken from the hardware. Static energy consumption estimation
enables energy-aware software development, without requiring hardware knowledge.

\end{abstract}

\section{Introduction}
\label{sec:introduction}

In embedded systems, low energy consumption is a very important
requirement. The software running on these systems has a profound effect on the
energy consumed. The design of software and algorithms, the programming
language and the compiler together with its optimization level all contribute
towards energy consumption of an application. Measuring such consumption,
however, requires hardware specific knowledge and instrumentation, making
such measurements challenging for software engineers.

Estimations of energy consumption of
programs are very useful to software engineers, so that they can understand the effect
of their code on the energy consumption of the final system, without the
need to instrument or even have the system. Accurate energy
consumption and timing analysis of programs involves analyzing low-level machine
code representations. However, programs are written in
high-level languages with rich abstraction mechanisms, and the relation between
the two is often blurred. For instance, optimizations such as dead code
elimination, various kinds of code motion, inlining and other clever loop
optimization techniques obfuscate the structure of the program and make the
resultant code difficult to analyze~\cite{Tice1998}.

In this paper, we develop a static analyzer that works on the intermediate
compiler representation of the program (LLVM IR). Our analysis is based on a
well-developed approach in which recursive equations (cost relations) are
extracted from a program, representing the cost of running the program in terms
of its input~\cite{DBLP:journals/cacm/Wegbreit75,
  Rosendahl89,AlbertAGP11a,Albert:2009:CRS:1594413.1594567}.
These cost relations are finally
converted to \emph{closed-form}, i.e. without recurrences, by means of a
solver. For example, we can analyze the following program.

{
\nocaptionrule
\begin{lstlisting}[language=C]
  void proc(int v[], int l) {
      for (int i = 0; i < l; i++)
        if(v[i] & 1)
          odd();
        else
          even();
  }
\end{lstlisting}
}

\noindent The following CRs are extracted from the program,

\begin{align*}
(a)&\phantom .\CR{proc}(l)=k_1+\CR{for}(l,0) &\text{ if }l\geq 0\\
(b)&\phantom .\CR{for}(l,i)=k_2         &\text{ if }i\geq l \wedge l\geq 0\\
(c)&\phantom .\CR{for}(l,i)=k_3+\CR{odd}() + \CR{for}(l,i+1) &\text{ if }i\leq l \wedge l\geq 0\\
(d)&\phantom .\CR{for}(l,i)=k_4+\CR{even}() + \CR{for}(l,i+1) &\text{ if }i\leq l \wedge l\geq 0\\
\end{align*}
where $l$ denotes the length of the array \texttt{v}, $i$ stands for the counter of the
loop and $C_{proc}$, $C_{odd}$ and $C_{even}$ approximate, respectively, the
costs of executing their corresponding methods. The constraints, denoted on the
right hand side of the relations, specify a condition that must be true for the
cost relation to be applicable. For instance, relation $(a)$
corresponds to the cost of executing \texttt{proc} with an array of length
greater than 0 (stated in the condition $l > 0$), where cost $k_1$ is
accumulated to the cost of executing the loop, given by $C_{for}$. Note that the
transition into $(c)$ and $(d)$ is non deterministic. The constants
$k_1,\ldots,k_4$ take different values depending on the cost model that one
adopts. In this paper, our cost model focuses on energy. These constants are
obtained from energy models created at the Instruction Set Architecture (ISA)
level~\cite{Kerrison13}. Such models have previously been applied to analysis at
the same level~\cite{NMHLFM08, isa-energy-lopstr13}, and in this paper we
propagate this up to the LLVM level.

Many modern compilers such as Clang or XCC are built using the LLVM framework. These
internally transform source programs into intermediate compiler representations,
which are more amenable to analysis than either source or machine level
programs. We show
how resource consumption analysis techniques can be adapted and applied to
programming languages targeting LLVM IR (such as C or XC \cite{xc}) by reusing some of the
existing machinery available in the compiler framework (for
instance LLVM analysis passes).
We show how cost relations  can be extracted from programs, such that these can
be solved using an existing solver \cite{AlbertAGP11a}. Specifically, we focus on
\emph{optimized LLVM IR}, that has been compiled with optimization levels
used in production software (i.e. \texttt{O2} or higher).

Time is a significant component of energy consumption, in that a program that
computes its result quicker will typically consume less energy by virtue of a
shorter run-time. However, the correlation between time and energy varies
between architectures, and is related to the complexity of the processor's pipeline~\cite{Pallister2013}.
For example, one of the target architectures for this
paper exhibits an approximately $2\times$ difference in energy depending on the
instructions that are executed, with a similar relationship for the number of
threads executed upon it~\cite{Kerrison13}. Analysis of system energy and not
just of execution time will therefore garner better information on the energy
characteristics of a program.

\begin{figure}[hb]
\centering
\includegraphics[width=2.5in]{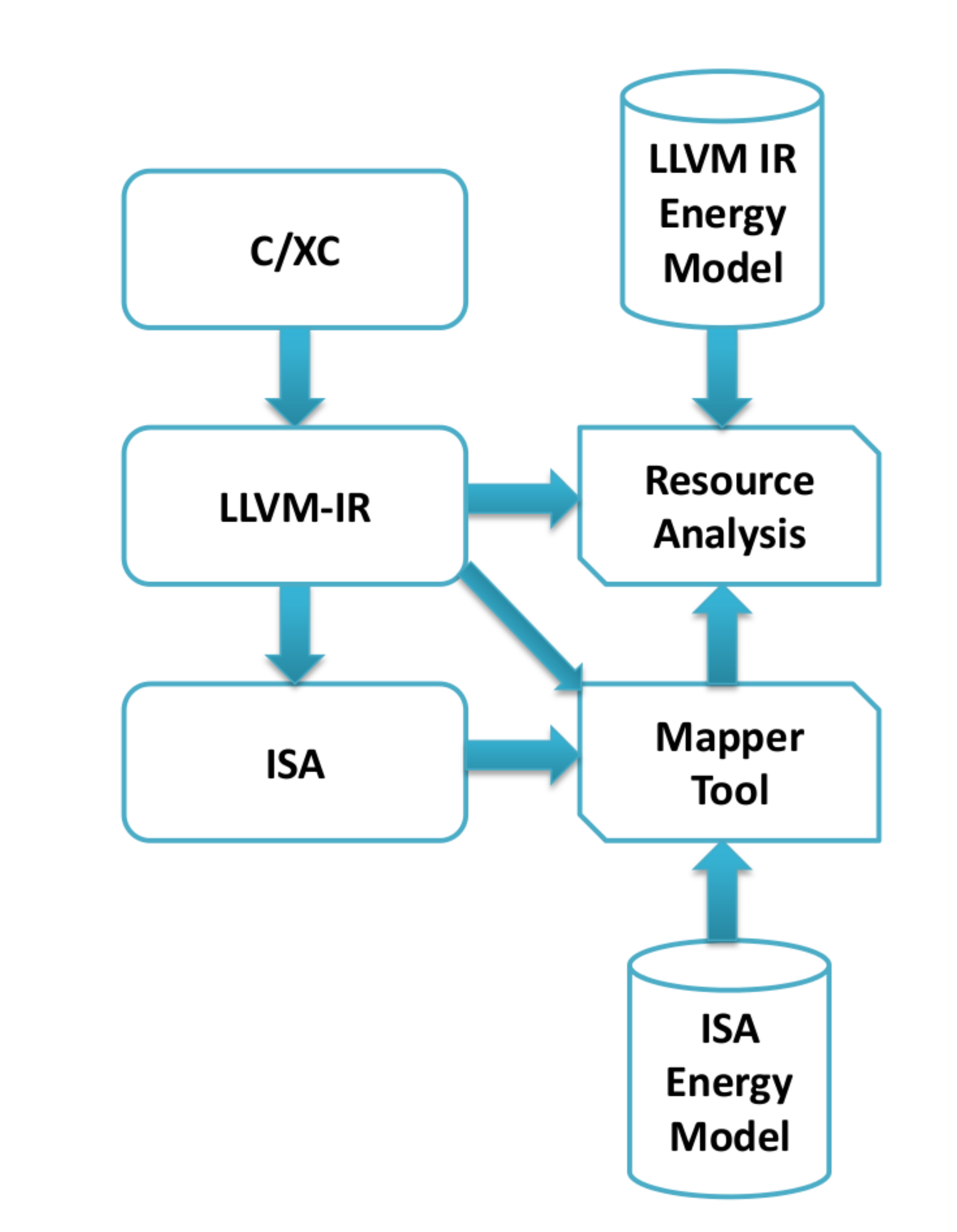}
\caption{\figlabel{llvmiranalysis}Illustration of the analysis toolchain.}
\end{figure}

Energy models can be constructed for a processor's instruction set, however this
information needs to be constructed, or propagated to a higher level program
representation in order to benefit our analysis mechanism. We propose two
different techniques (Section 4), for assigning energy to a
higher level program representation (LLVM IR).
We first propose a mechanism for \emph{mapping} program segments at ISA level
to program segments at LLVM IR level. Using this mapping, we can perform a multi
level program analysis where we consider the LLVM IR for the structure and semantics
of the program and the ISA instructions for the physical effect on the
hardware. We also propose an alternative technique, of determining
the instruction energy model directly at the LLVM IR level. This is based on
empirical data and domain knowledge of the compiler backend and underlying processor.

This paper focuses on static analysis of code for processors that are embedded
or deeply embedded. Such processors do not typically feature cache hierarchies.
They have small amounts of static-RAM and possibly flash memory available to
them. This constrains the application space, but the motivation for analysing
software that targets these processors is greater, because these types of
embedded systems often have the strictest energy consumption requirements.

The analysis toolchain is illustrated in \figref{llvmiranalysis}. The static
resource consumption analysis mechanism is described in Section 3. Parts of this mechanism
perform a symbolic execution of LLVM IR, which is described in Section 2. The
techniques described are built into a tool, which can be integrated into the
build process and statically estimates the energy consumption of an embedded
program (and its constituent parts, such as procedures and functions) as a
function on several parameters of the input data. Our approach is validated in
Section 5 on a number of embedded systems benchmarks, on both xCORE and Cortex-M
platforms. Finally, we describe related work in Section 6 and conclude in
Section 7.

\section{Structure and interpretation of LLVM IR}
\seclabel{statLLVMIR}
In this section we describe the core language and an important technique we
utilize in the resource consumption analysis mechanism (\secref{resourceanalysis}), which
infers energy formulae given an LLVM IR program.

\subsection{The LLVM IR language}
LLVM IR is a Static Single Assignment (SSA) based representation. This is used
in a number of compilers, and is designed to represent high-level
languages. For presentation purposes, we first formalize a simple
calculus of LLVM IR, based on the following syntax:
\begin{align*}
inst &= \br~p~\aBlock_1~\aBlock_2 \quad \text{(conditional branch)}\\
     &\mid x=\op~a_1 .. a_n  \quad \text{(generic op., no side-effects)}\\
     &\mid x=\phi~\tuple{\aBlock_1}{x_1} .. \tuple{\aBlock_n}{x_n} \quad \text{(phi nodes)}\\
     &\mid x=\call~f~a_1~..~a_n\\
     &\mid x=\memload \quad \text{(dynamic memory load)}\\
     &\mid \memstore \quad \text{(dynamic memory store)}\\
     &\mid \ret~a
\end{align*}
We use metavariable names $p,f,a,x$ to describe predicates, function names,
generic arguments and variables respectively. The concrete semantics of the
instructions are modeled on the actual LLVM IR
semantics \cite{Zhao:2012:FLI:2103656.2103709}. Instruction $\op$ represents any
side effect free operation such as \texttt{icmp} or \texttt{add} in LLVM. The
$\phi$ instruction takes a
list of pairs as arguments, with one pair for each predecessor basic block of
the current block. Each pair contains a reference to the predecessor block
together with the variable that is propagated to the current block. The only
place where a $\phi$ instruction can appear is in the beginning of a basic
block. Two interesting instructions are $\memload$ and $\memstore$. These represent any
dynamic memory load and store operation respectively. For instance,
\texttt{getelementptr} and \texttt{load} are some examples of instructions
represented by $\memload$. These instructions typically compute pointers dynamically and load
data from memory. In our abstract semantics of LLVM IR, we therefore treat
variables assigned with values dynamically loaded from memory as unknown
(denoted `$?$').

LLVM IR instructions are arranged in \emph{basic blocks}, labeled with
a unique name. A basic block \aBlock over a CFG is a maximal sequence of
instructions, $\inst_1$ through $\inst_n$, such that all instructions
up to $\inst_{n-1}$ are not branch or return instructions and $\inst_n$ is \br
or \ret. The $\phi$ instructions always appear as the first instructions in a
block, as a block can have multiple in-edges. All \call instructions are assumed
to eventually return.

\subsection{Symbolic evaluation of LLVM IR variables}\seclabel{seval}
At the core of our resource consumption analysis mechanism of LLVM IR is a symbolic
evaluation function \seval. Given a block of code \aBlock, and a variable \aVar,
$\seval(\aBlock,\aVar)$ symbolically executes this block, producing a slice
\cite{slicing} of the block with respect to \aVar. During this
static analysis phase, we apply an abstract semantics of LLVM IR, which
abstracts away dynamic memory reads and writes i.e., \memload and
\memstore. This has the effect of producing simple expressions, which can be
handled by the PUBS solver. The algorithm proceeds by starting at the last
assignment of \aVar in the block, and evaluates the assigned expression using
this semantics, recursively evaluating all its dependencies until an expression
or variable outside the block is reached. For example, given the following snippet:

\begin{llvmcode}
  LoopIncrement:
    %postinc = add i32 %i.0, 1
    %exitcond = icmp eq i32 %postinc, %1
    br i1 %exitcond, label %return, label %LoopBody
\end{llvmcode}
$\seval(\ldots,$\llvminline{\%exitcond}$)$ is $($\llvminline{\%i.0}$ + 1) ==
$\llvminline{\%1}, while in the following snippet

\begin{llvmcode}
  iftrue2:
    call void @odd()
    br label %LoopIncrement
\end{llvmcode}
$\seval(\ldots$,\llvminline{\%i.0}$)$ would evaluate to \llvminline{\%i.0},
because there are no assignments to \llvminline{\%i.0}.

\section{Resource Consumption Analysis for LLVM IR}\seclabel{resourceanalysis}
The techniques described here are used to infer cost relations
\cite{AlbertAGP11a}. Cost relations are recursively defined and closely follow the flow of the
program. What we actually want to infer is a closed form formula modeling the
cost, which is parametric to any relevant input arguments to the program, which
requires solving using a cost relation solver. These solvers typically work with
simplified control flow graph structures, and therefore we must first perform some
simplifications on the control flow graphs, as described in
\secref{transformations}. The analysis then infers block arguments by using
symbolic evaluation as described in \secref{seval}.

\subsection{Inferring block arguments}
Block arguments characterize the input data, which flows into the block, and is
either consumed (killed) or propagated to another block or function. Unfortunately, solving multi-variate cost relations
and recurrence relations automatically is still an open problem, and the fewer
arguments each relation has, the easier it is to solve these. For this reason,
we designed an analysis algorithm to minimize the block arguments before
inferring the cost relations.

The algorithm for inferring block arguments is a data flow analysis
algorithm. We use a standard means to describe this algorithm, as in
\cite{Nielson99}.
We define a data flow analysis function \Gen,
which, given a basic block, returns the variables of interest in that block:
\begin{equation*}
\Gen(\aBlock)=\Gen_{blk}(\aBlock)\cup \Gen_{fn}(\aBlock)
\end{equation*}
The function $\Gen_{blk}$ returns the input arguments that affect the branching in a block
\aBlock, composed of instructions $\inst_1$
through $\inst_n$, and $\Gen_{fn}$ returns the variables that affect the input to any
external calls in the block. $\Gen_{blk}$ is defined as follows:
\begin{align*}
\Gen_{blk}(\aBlock)&=
\begin{cases}
 \Ref(\seval(\aBlock,p)) &\text{if }\inst_n=[\br~p~..]\\
 \emptyset & \text{otherwise}
\end{cases}\\
\end{align*}
The function \Ref returns all variables referred to in the symbolically
evaluated expression given as argument, for example $\Ref(x>(y+3))$ returns
$\{x,y\}$. We also define function $\Gen_{fn}$. This returns all the input
arguments that affect the parameters given to the function, and is defined as:
\begin{equation*}
\Gen_{fn}(\aBlock)=
\end{equation*}
\begin{equation*}
\bigcup\limits_{k=1}^{n}
\begin{cases}
\bigcup\limits_{i=1}^{m}\Ref(\seval(\aBlock,a_i))&\text{ if } \inst_k \text{ is
}[x=\call~f~a_1~..~a_m] \\
\emptyset & \text{otherwise}
\end{cases}
\end{equation*}

\noindent
The data flow analysis function \Kill is defined as:
\begin{equation*}
\Kill(\aBlock)=\bigcup\limits_{k=1}^n
\begin{cases}
\{x\} &\text{if }\inst_k \text{ is } x=\call~\ldots\\
\{x\} &\text{if }\inst_k \text{ is } x=\op~\ldots\\
\{x\} &\text{if }\inst_k \text{ is } x=\memload~\ldots\\
\emptyset & \text{otherwise}
\end{cases}
\end{equation*}

Finally, we combine $\Gen$ and $\Kill$ by utilizing a transfer function, which
is inlined into $\args_{in}$ and $\args_{out}$. These compute the relevant block
arguments utilized by the resource consumption analysis. $\args_{in}(\aBlock)$ is defined as
the function's arguments if \aBlock is the function's first block. In all other
cases, $\args_{in}$ and $\args_{out}$ are defined as:
\begin{align*}
\args_{out}(\aBlock) &=
  \kern-1em\bigcup\limits_{\aBlock'\in\flownext(\aBlock)}\kern-1em phimap_{\tuple{\aBlock}{\aBlock'}}(\args_{in}(\aBlock'))\\
\args_{in}(\aBlock)&=(\args_{out}(\aBlock)-\Kill(\aBlock)) \cup \Gen(\aBlock)
\end{align*}
where $phimap$ maps variables between adjacent blocks \aBlock and $\aBlock'$
based on the $\phi$ instructions in $\aBlock'$.

Functions $\args_{in}$ and $\args_{out}$ are recomputed until their
least fixpoint is found. Finally, the block arguments are found in $\args_{in}$.
The analysis explained in this section is closely related to live variable
analysis. A crucial difference, however, is in the function \Gen. In our case,
this returns a smaller subset of variables than live variable analysis i.e.,
only the ones that may affect control flow.

\subsection{Generating and solving cost relations}
\seclabel{CRs}
In order to generate cost relations we
need to characterize the energy exerted by executing the instructions in a
single block. We also need to model the continuations of each
block. Continuations, expressed as calls to other cost relations, arise from
either branching at the end of a block, or from function calls in the middle of a
block. For instance, consider the following LLVM IR block:

\begin{llvmcode}
  LoopIncrement:
    %postinc = add i32 %i.0, 1
    %exitcond = icmp eq i32 %postinc, %1
    br i1 %exitcond, label %return, label %LoopBody
\end{llvmcode}

\noindent This would translate to the following relation:
\begin{align*}
\CR{LI}(i) &=  C_0 + \CR{ret}(i+1)         &\text{if }i+1 = a_1~ \\
\CR{LI}(i) &=  C_1 + \CR{LB}(i+1)         &\text{if }i+1 \neq a_1 \text{,}\\
\end{align*}
where \CR{LI}, \CR{ret} and \CR{LB} characterize the energy exerted when running the blocks
\texttt{LoopIncrement}, \texttt{return} and \texttt{LoopBody}
respectively. We therefore refer to \CR{ret} and \CR{LB} as continuations of
\CR{LI}. Expressing these calls to other cost relations involves evaluating
their arguments, which cannot be done without evaluating the
program. Instead, by symbolically executing the block, we can express the
arguments of the continuation in terms of the input arguments to the block. In
order to do so, we perform symbolic evaluation using the function \seval.

The cost relations, extracted from recursive programs using the techniques
discussed in this section, can be automatically transformed to closed form by PUBS
\cite{AlbertAGP11a}.
PUBS infers closed form solutions recursively, starting with the inner-most
relations, using various techniques such as computing ranking functions and loop
invariants. The results of the intermediate steps are then mathematically composed to
solve the whole set of given cost relations.

There are cases where the optimized program structures produced by LLVM based
compilers prevent the cost relation solvers from finding unique \emph{cover
points} in the structure of the cost relations. In order to solve this problem,
we need to perform transformations to the call graph upon which we construct our
cost relations. This is described in the next section.

\subsection{Transformations for control flow graphs}\seclabel{transformations}
After compilation, nested loop program structures are mangled
by compiler optimizations. When the resulting Control Flow Graph (CFG) is directly
used to produce CRs, it is usually not possible to infer closed form
solutions. For instance PUBS \cite{AlbertAGP11a} cannot handle complex CFGs, and
therefore in order to analyze programs with nested loops, the
CFG needs to be simplified. The simplification is actually done at an early
stage in the analysis, right after generating an initial CFG,
using the following steps:

\begin{enumerate}
  \item Identify a loop's CFG, A, that has nested loops.
  \item Identify the sub-CFG, B, of A corresponding to the inner loop.
  \item Extract B out of A, so that B is a separate CFG. This can be thought
    of as a new function with multiple return points. Hence B's exit edges are
    removed.
  \item In A, in the place where B used to be, keep the continuation to B.
    Append a continuation to B's exit targets to B's caller in A.
\end{enumerate}

In order to perform the first two steps, we need to identify the loops in the
CFG. While LLVM has specific passes to do so, we had better success when using
the algorithm described in \cite{Wei2007}. As an example, we show how these
steps can be used to transform the CFG of a simple insertion sort, as shown in
Listing~\ref{fig:insertion}. The original CFG of this program, when compiled using
\texttt{clang} with optimization level \texttt{O2} is shown in
\figref{insertioncfgbefore} (left). In this CFG, the nested loops are identified, which
also involves identifying their corresponding entries, re-entries, exit and loop
headers. Here, blocks \texttt{bb1}, \texttt{bb2} and
\texttt{.backedge} form the inner loop. These blocks are hoisted and the exit
edge from \texttt{.backedge} (dotted) is eliminated. Instead, \texttt{.loopexit}
is then called after \texttt{bb1} ``returns'' (\figref{insertioncfgafter}).

The CFG simplifications described in this section preserve the same order of
operations when applied to an existing CFG compiled from typical \texttt{while}
or \texttt{for} using \texttt{clang} or \texttt{xcc}. This means that the program
called in the left-side of \figref{insertioncfgafter} will consume as much
energy as the program in the right-side of~\figref{insertioncfgbefore}. The only limitation of this approach is
when an induction variable of an outer loop is modified in an inner
loop. In this case the transformation cannot occur, however we have not
encountered real benchmarks where this takes place.

In order to verify the transformation with respect to energy, let us consider a
typical \texttt{while} or \texttt{for} loop and show that the same sequence of
blocks is called after the transformation takes place. We can assume that such a
loop has a single header, but may have multiple exits or reentries and induction
variables of the outer loops are not modified in the inner loops. After the
transformation takes place on a nested loop structure (B inside A), B is still
called from A, however B's exit edges are now removed. The target of B's
exit edges will still be called after B completes. This is because we have
appended a continuation in A to this target, in Step 4. Hence all blocks will be
called in the same sequence. The argument above can be inductively applied to
loops with arbitrary nesting levels.

\begin{lstlisting}[language=C,float,caption={This insertion sort demonstrates that certain classes of programs require further analysis or transformation.},label=fig:insertion]
  void sort(int numbers[], int size) {
    int i=size, j, temp;
    while(j = i--)
      while(j--)
        if(numbers[j] > numbers[i]) {
          temp = numbers[i];
          numbers[i] = numbers[j];
          numbers[j] = temp;
        }
  }
\end{lstlisting}

\begin{figure}
\begin{center}
\includegraphics[height=10cm,clip,trim=1.25cm 1.25cm 1.25cm 1.25cm]{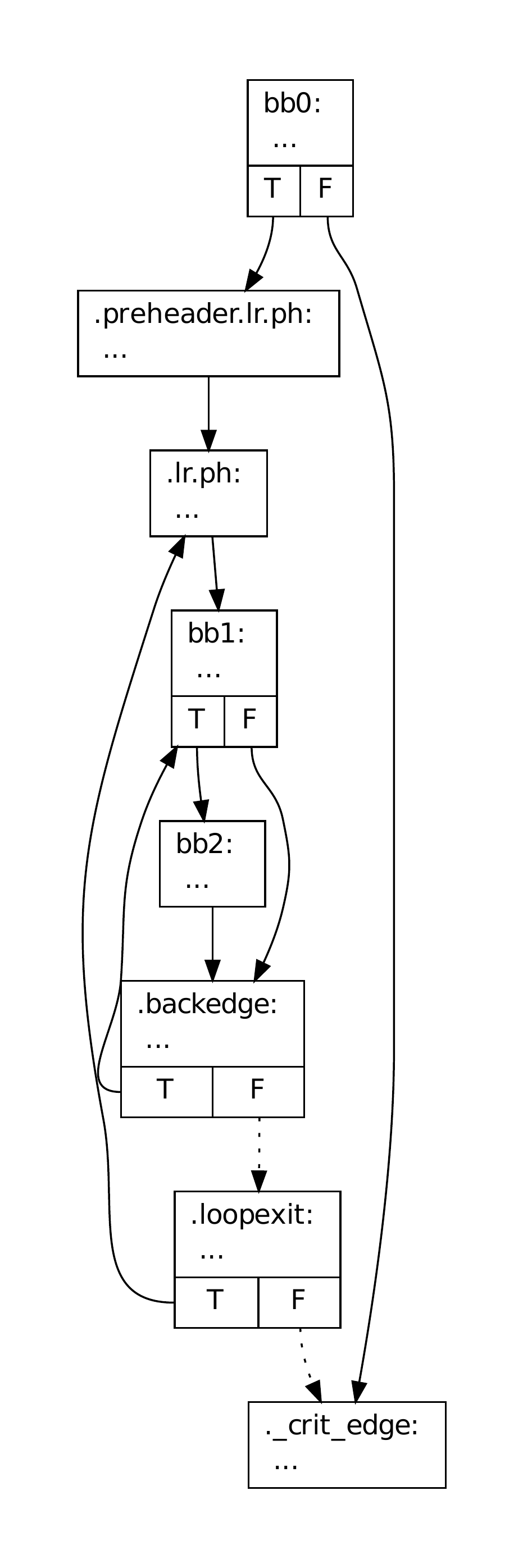}
\includegraphics[height=10cm,clip,trim=1.25cm 1.25cm 1.25cm 1.25cm]{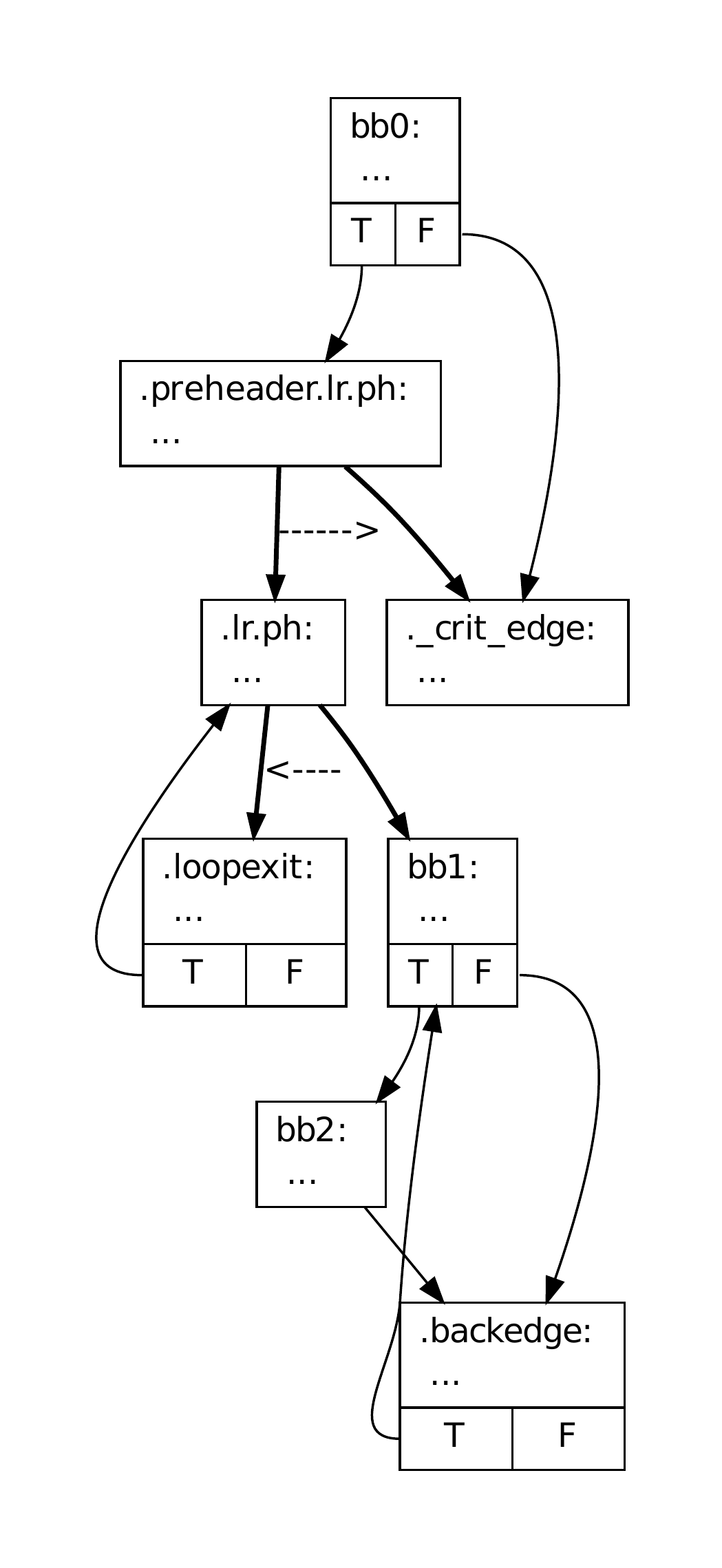}
\end{center}
\caption{\figlabel{insertioncfgbefore}\figlabel{insertioncfgafter}
CFG of an insertion sort compiled using \texttt{clang} with optimization
  level \texttt{O2} before (left) and after simplification (right).}
\end{figure}

\section{Computing energy cost of LLVM IR blocks}

The intermediate representation used by LLVM is architecture independent. Any
given LLVM IR sequence can be passed to one of many different backends,
including ISAs~\cite{LLVM-backend}.  The exact implementation of the ISA
determines the energy consumed by each instruction that is executed.  Thus, the
conversion to machine code, together with the processor implementation, affects
the energy consumption of an instruction at the LLVM IR level.

For static analysis of LLVM IR to produce useful energy formulae for programs, a
method of assigning an energy cost to an LLVM IR segment must be used. Two
possible methods are demonstrated in this paper:
\begin{enumerate}
  \item \emph{ISA energy model w/mapping}. LLVM IR is mapped to its corresponding
    ISA instructions and the energy cost is obtained from the ISA level cost
    model. The advantage is that it is simpler to characterize at ISA level, however this requires an additional step to correlate
    LLVM with ISA instructions.
  \item \emph{LLVM energy model}. Attributing costs directly to LLVM IR removes
    the need for a mapping. However, it necessarily simplifies the energy
    consumption characteristics, reducing accuracy.
\end{enumerate}

In principle, both methods can be explored for both architectures. This paper
utilizes an ISA level model for the XMOS processor. The Cortex-M is
modeled at the LLVM IR level directly.

\subsection{XMOS XS1-L ISA level modeling}
\label{sec:xs1model}

The aim of ISA level modeling is to associate machine instructions with an
energy cost. To achieve this, energy consumption samples must be collected and
an appropriate representation of the underlying hardware must be used as a basis
for the model. A single-threaded model, such as that defined by
Tiwari~\cite{Tiwari-embedded-1994} and expressed in Equation~\ref{eq:tiwari},
describes the energy of a sequence of instructions, or program.
\begin{align}
E_{\text{prog}} =
        \sum_{i \in \text{ISA}} \left( B_i N_i \right)
        + \sum_{i,j \in \text{ISA}} \left(O_{i,j} N_{i,j} \right)
        + \sum_{k \in \text{ext}} E_k
\label{eq:tiwari}
\end{align}
The program's energy, $E_{\text{prog}}$, is first formed from the base cost,
$B_i$ of all instructions, $i$, in the ISA, multiplied by the occurrences,
$N_i$, of each instruction. For each transition in a sequence of instructions,
the overhead, $O_{i,j}$, of switching from instruction $i$ to instruction $j$,
multiplied by the number of times the combination $i,j$ occurs, $N_{i,j}$.
Finally, for a set of $k$ external effects, the cost of each of these effects,
$E_k$ is added. For example, these external effects may represent the cache and
memory costs, based on the cache hit rate statistics of the program.

The XS1-L architecture implements multi-threading in a hardware pipeline. Even
for single-threaded programs, we need to consider the behavior of this
multi-threaded pipeline. The power of individual instructions varies by up to
$2\times$, with multi-threading introducing up to a $1.6\times$ increase with a $4\times$ performance
boost.  This means execution time and energy are related in a more complex way
than a simpler single-threaded architecture. The model for the XS1-L is built
upon existing work of~\cite{Tiwari-Instruction-level-1996}~and the more
detailed~\cite{Steinke2001}, which obtain model data through the energy
measurement of specific instruction sequences, and create a representation of
some of the processor's internal structure in the model equations. A full
description of the XS1-L's energy characteristics and the model is given
in~\cite{Kerrison13}.

To extend a Tiwari style approach to model the XS1-L processor, two new
characteristics must be accounted for: idle time and concurrency. The XS1
ISA has a number of event-driven instructions, which can result in the processor
executing no instructions for a period of time, until the event occurs.
Furthermore, the multi-threaded pipeline permits only one instruction from a
given thread to be present in the pipeline at any one time. These changes are
expressed in Equation~\ref{eq:model}. Here, the energy exerted
by running a program depends on a base power, $P_\text{base}$, which represents
the energy cost when no instructions are executed, multiplied by the number of idle periods,
$N_\text{idle}$. The clock period of the processor, $T_\text{clk}$ is also
introduced, to allow different clock speeds to be considered. The
inter-instruction overhead, previously described in Equation~\ref{eq:tiwari} as
$O_{i,j}$, is generalized to a constant overhead, $O$, due to the
unpredictability of instruction interaction between threads. For each
instruction, the base cost is added to the instruction cost, $P_i$, which is
scaled by the overhead and an additional scaling factor based on the number of
active threads, $M_t$. This is multiplied by the number of occurrences of this
instruction at $t$ threads, $N_{i,t}$ and the clock period, $T_\text{clk}$. This
is done for the varying number of threads, $t$ that may be active in the program
over its lifetime.
\begin{align}
\nonumber E_{\text{prog}} &=
        P_{\text{base}}N_{\text{idle}}T_{\text{clk}} \\
     &\quad + \sum_{t=1}^{N_{t}}{\sum_{i \in \text{ISA}}{\left(\left(M_{t}P_{i}O
        + P_{\text{base}}\right) \times \left(N_{i,t} T_{\text{clk}}\right)\right) }}
\label{eq:model}
\end{align}
The multi-threaded ISA level model for the XS1-L requires that for each level of
concurrency, $t$, the number of instructions executed at that level should be
known, or estimated. If a single threaded program is run on its own on the XS1-L
and there are no idle periods, then Equation~\ref{eq:model} simplifies to
Equation~\ref{eq:model1thread}, where the idle accounting is removed, and only
the first threading level, $t=1$, is considered.
\begin{align}
E_{\text{prog}} = \sum_{i \in \text{ISA}}{\left(\left(M_{1}P_{i}O
    + P_{\text{base}}\right) \times \left(N_{i} T_{\text{clk}}\right)\right) }
\label{eq:model1thread}
\end{align}

The current analysis effort focuses upon single threaded experiments, thus
Equation~\ref{eq:model1thread} can be used. Multi-threaded analysis is proposed
as future work in \secref{conc_future}. Temperature variation in the device is
not captured in this model, however prolonged testing of the target hardware
showed no significant temperature changes or associated affects that would
influence the single-threaded tests performed in this work.

\subsection{XMOS LLVM IR energy characterization by mapping}
\label{sec:mapping}
To enable the analysis at the LLVM IR level we need a mechanism to propagate
the existing energy model at the ISA level up to the LLVM level. The mapping
technique described in this section creates a fine grained
mapping between segments of ISA instructions to LLVM IR instructions, in order
to enable the energy characterization of each LLVM IR instruction in a program.
A full description of the mapping techniques is given in~\cite{Georgiou14}.

Our mapping technique leverages the existing debug mechanism in the XMOS compiler
toolchain. This mechanism is originally meant to
facilitate the debugging process of an application, particularly when stepping
through a program line by line.
During the lowering phase of the compilation process, the LLVM
IR code is transformed to the specific ISA code by the
backend. The debug information (DI) is also stored alongside with the ISA
code using the DWARF standard~\cite{DWARF:2013:Online}, a standardized
debugging data format used by many compilers and debuggers to support source level
debugging.
By tracking this information we can extract an n:m relationship between the two levels, because one
source code instruction can be related to many different sequences LLVM IR instructions and therefore
many different sequences of ISA instructions. Because this n:m relation complicates static
analysis, there is a need for a more fine grained mapping.

To address this issue, we created an LLVM pass that traverses the LLVM IR and
replaces the \emph{Source Location Information} with LLVM IR location information,
right after all the optimization passes and just before emitting the ISA code.
In this way, we can extract a 1:m relationship between the mapping of LLVM
IR instructions and ISA instructions.
Also, by doing it after the LLVM optimizations
passes the optimized LLVM IR is closer to the ISA code than the unoptimized one,
which will go through a series of transformations.
There are optimizations that happen during the lowering phase,
such as peephole optimizations and some late target specific optimizations
that can affect the mapping. However, the effect of these optimizations on the
structure of the code is not as profound as those
applied to LLVM IR. After a mapping is extracted for a particular program,
the associated energy values for the ISA instructions corresponding
to a specific LLVM IR instruction are aggregated and then associated with the LLVM
IR instruction, and finally to every LLVM IR block.

Although we use the XMOS tool-chain for the mapper tool, the approach is generic
and
transferable, due to the use of the common LLVM optimizer and code generator,
and the use of the DWARF standardized debugging data format, used by many
compilers and debuggers to support source layer debugging.

\subsection{LLVM IR energy model for ARM}
An energy model for ARM Cortex-M series is applied directly at the LLVM IR
level, based upon empirical energy measurement data, and knowledge of both the
processor architecture and the compiler backend. The Cortex-M3 model is for the
most part a simplification of the Tiwari model~\cite{Tiwari-embedded-1994},
applied at the LLVM IR level.  The processor does not does not feature a cache,
so it is not necessary to model cache misses as external effects. The effect of
the switching cost between instructions is approximated into the actual
instruction cost, rather than assigning a unique overhead for each instruction
pairing.

Through analysis of energy measurements for a large set of the target ISA
instructions, it was found that LLVM IR instructions can be segmented into four
groups: memory, $M$, program flow, $B$, division, $D$, and all other
instructions, $G$. The LLVM IR syntax described in \secref{statLLVMIR} can be
related to these groupings. In particular, $\br$, $\call$ and $\ret$ can be
combined into group $B$; $\memload$ and $\memstore$ are members of $M$; the
subset of $\op$ relating to division make up group $D$; and finally, $\phi$ and
all remembering members of $\op$ form group $G$.

This yields a model equation that accumulates the energy of a program based on
the number of instructions executed from each group. Equation~\ref{eq:m3model}
considers each group, which is assigned an energy cost, which combined give
the total program energy, $E_\text{prog}$, where $E_i$ is the energy cost of a
single instruction in group $i$, and $N_i$ is the number of instructions executed in
that group.
\begin{align}
E_{\text{prog}} =
    \sum_{i \in \{M,B,D,G\} }{
        E_i N_i
    }
\label{eq:m3model}
\end{align}
In addition, there are a number of other factors that affect energy, due to the
relation between the LLVM IR and the ISA:

\begin{enumerate}

\item \textbf{Variadic arguments.}  LLVM has instructions with variadic
arguments. Typically, the number of arguments in the instruction affects the
energy consumed in a linear manner.
\vspace{-1mm}
\item \textbf{Data types.} LLVM operations \op can be performed on values of
different data types. If the data type is larger than 32 bits, or floating
point, this will  translate into a larger number of ISA instructions on a
Cortex-M with no floating point unit.

\item \textbf{Predicated instructions.} The Cortex-M processor is capable of
executing predicated instruction sequences. In some cases, short LLVM IR blocks
originating from ternary expressions in the original source code are directly
translated to a number of predicated instructions in the ARM ISA. Therefore, the
number of ISA instructions generated could be less than the instructions in LLVM
IR, and the static analysis over-approximates the energy consumption of these
blocks.
\end{enumerate}

Factors (1) and (2) can be accounted for by parameterizing the LLVM IR energy
model. For instance, consider the following call instruction:
\begin{llvmcode}[numbers=none]
  %6 = call i32 @min(i32 %boptmp88, i32 %boptmp96)
\end{llvmcode}
This translates to a single branch instruction in the ARM ISA, with surrounding
register moves to ensure the correct calling convention:

{\nocaptionrule
\begin{lstlisting}[morekeywords={mov,bl},morecomment={[l]\#}]
  mov r0, r4    # move arg1 into r0
  mov r1, r5    # move arg2 into r1
  bl min        # call min
  mov r4, r0    # move the result into r4
\end{lstlisting}}
As we can see, the energy consumed by an LLVM \call instruction is
parametric in the number and types of the arguments and return value.

\section{Experimental Evaluation}

We have selected a series of benchmarks of core algorithmic functions,
particularly from the BEEBS~\cite{DBLP:journals/corr/PallisterHB13} and MDH WCET
benchmark~\cite{Gustafsson2010} suites. These are collections of open source
benchmarks for deeply embedded systems, where the activities performed in these
benchmarks are typical of such systems. Analysing benchmarks of this size and
with their particular characteristics is therefore a good means of evaluating
our analysis technique in order to demonstrate its usefulness within the
embedded systems software space. The benchmarks are single threaded, reflecting
the scope of the analysis performed in this paper. Minimal modifications were
made to allow integration into our test harness.  Table~\ref{fig:bench-Char}
summarizes the characteristics of the benchmarks and the meaning of the last 5
columns is as follows: (L) contains loops, (NL) contains nested loops, (A) uses
arrays and/or matrices, (B) contains bitwise operations, (C) contains loops with
complex control flow predicates.

In order to show that our techniques are applicable to multiple languages and platforms,
we have ported some of the benchmarks from C to XC. Porting C code to XC typically does
not involve rewriting, since the syntax is very similar and they both use the
same preprocessors. However, since XC does not provide pointers some changes
need to be made to the benchmarks during the porting process. For the benchmarks
that run on the xCORE, we have used the XC compiler, version 13. For Cortex-M
benchmarks we have used Clang version 3.5. We proceed by describing the
benchmarks. In both cases, the benchmarks are compiled under optimization level
\texttt{O2}.

\begin{table}
\centering
\nocaptionrule
\begin{tabular}{l*{6}{c}r}
\toprule
Benchmark         & L & NL & A & B & C\\
\midrule
base64            & \texttimes &  & \texttimes & \texttimes &\\
mac               & \texttimes &   & \texttimes & &\\
levenshtein       & \texttimes & \texttimes   & \texttimes & \texttimes &\\
insertion sort    & \texttimes & \texttimes  & \texttimes & &\\
matrix multiply   & \texttimes & \texttimes  & \texttimes & &\\
jpegdct           & \texttimes & \texttimes & \texttimes & \texttimes & \texttimes\\
\bottomrule
\end{tabular}
\caption{Benchmark Characteristics.}
\label{fig:bench-Char}
\end{table}

\paragraph{Insertion sort.}
The code of the main function is shown in \figref{insertion}. The
energy exerted by the insertion sort partly depends on how many swaps need to take
place, and this is dependent on the actual data present inside the array. Since
PUBS infers a formula representing an upper bound of the closed form solution,
we will be measuring the energy consumed in sorting a reverse-ordered list, and
comparing this to the statically 
inferred formula. Note that the number of iterations in the inner loop depends
on an induction variable in the outer loop. This benchmark is parameterized by
the length of the list to be sorted, $P$.

\paragraph{Matrix multiply (BEEBS/MDH WCET).} We slightly
modified this so that it can work with matrices of various sizes. The matrices are all square, of size $P$.

\paragraph{Base64 encode.}
Computes the base64 encoding\footnote{Posted by user2859193 on
  stackoverflow.com/questions/342409} as a string, given an input string of length $P$.

\paragraph{MAC (MDH WCET).} Dot product of two vectors together with sum of
squares. Parameterized by the length of the vectors, $P$.

\paragraph{Jpegdct (MDH WCET).} Performs a JPEG discrete cosine transform. Taken
from the MDH WCET benchmark suite. This benchmark is not parameterized.

\paragraph{Levenshtein distance (BEEBS).} Computes the minimum number of edits to
change one string into another. The lengths of the two strings are
parameterized with the variables $A$ and $B$.

\subsection{Experimental Setup}
For both ARM and XMOS platforms, power measurement data is collected by using
instrumented power supplies, a power sense IC and an embedded
system running control and data collection software. The implementations
differ, but are structurally very similar. Both of these periodically calculate the
power using Equation~\ref{eq:powercalc} during a test run by sampling the
voltage on either side of a shunt resistor ($V_\text{bus}$ and $V_\text{shunt}$)
to determine the supplied current.

\begin{equation}
P = I \times V_\text{bus} \quad \text{where} \quad
I = \frac{V_\text{bus} - V_\text{shunt}}{R_\text{shunt}}
\label{eq:powercalc}
\end{equation}

For the Cortex-M processor, the measurements are taken on an ST Microelectronics
STM32VLDISCOVERY board while for the xCORE, a custom XMOS board with an
XS1-L based XS1-U16A chip is used.

\subsection{Results}

\begin{figure}
  \centering
  \includegraphics[width=\linewidth]{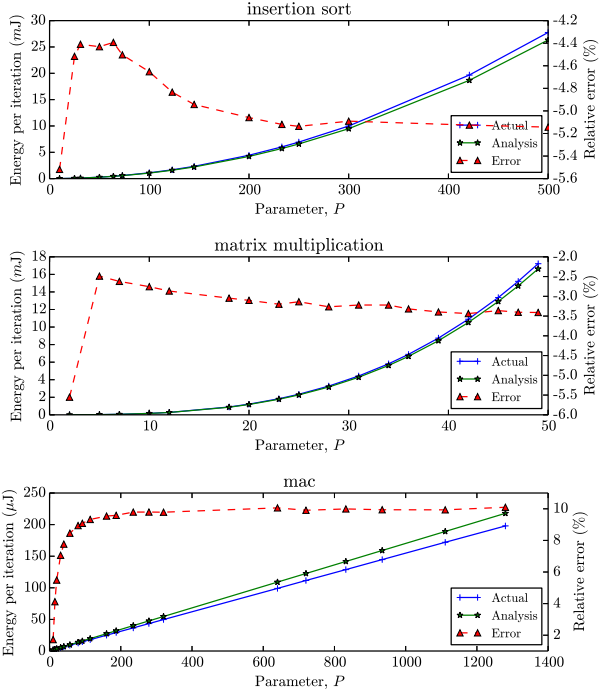}
  \caption{The measurement results and static analysis for the XMOS processor.}
  \label{fig:xmos_results}
\end{figure}
\begin{figure}
  \centering
  \includegraphics[width=\linewidth]{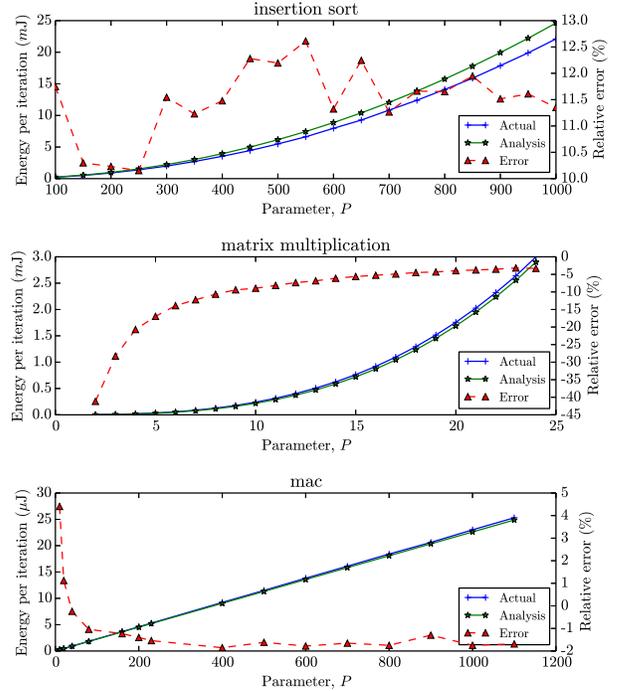}
  \caption{The measurement results and static analysis for the Cortex-M processor.}
  \label{fig:arm_results}
\end{figure}

\begin{table*}
  \centering
  \begin{tabular}{lllrr}
  \toprule
  Benchmark\hspace{-1mm}$\phantom.^*$ & \multicolumn{2}{c}{Formulae} & \multicolumn{2}{c}{Final error (\%)} \\
                  & ARM (nJ) &XMOS (nJ) &  ARM & XMOS \\
  \midrule
  base64          & $158 + 94 \cdot \left\lfloor \frac{P-1}{3} \right\rfloor$ & $1270 + 734 \cdot \left\lfloor \frac{P-1}{3} \right\rfloor$   & 28.0                                    & 1.1                                   \\
  mac             & $23P + 14$                                                & $133 P+192$                                                   & -1.7                                    & 10.1                                  \\
  levenshtein     & $47AB+14A+31B+44$                                         &
  $559AB + 78A+571 + \max(225B,180B+213)$                         & 7.0                                     & 0.4                                   \\
  insertion sort  & $25P^2 + 11P + 7.1$                                       & $105 P^2+30 P+75$                                             & 11.1                                    & 3.0                                   \\
  matrix multiply & $20P^3 + 13P^2 + 97P + 84$                                & $144 P^3+200 P^2+77 P+332$                                    & -3.3                                    & -3.4                                  \\
  jpegdct         & 54 mJ\hspace{-1mm}$\phantom .^\ddagger$                   & 463 mJ\hspace{-1mm}$\phantom .^\ddagger$                     & 8.5                                     & 2.6                                   \\
  \midrule
  \textit{Mean relative error} & && 9.9 & 3.4 \\
  \bottomrule
  \end{tabular}
  \nocaptionrule
  \caption{Formulae and error values for each benchmark.}
  \vspace{2mm}
  $\phantom .^\ddagger$ This benchmark was not parameteric, thus is not parameterised. \\
  \raggedleft \small\hspace{-1mm}$\phantom .^*$ Benchmark sources are available
  from \url{https://github.com/mageec/beebs}\\
  \label{tab:table_results}
  \vspace{-1em}
\end{table*}

\begin{figure}
\begin{minipage}[b]{0.53\linewidth}
{\nocaptionrule
\begin{lstlisting}[language=c,gobble=4]
    void function(int A, int B) {
      int i;

      if(A < B)
        for(i = 2*A; i >= 0; i--)
          ...
      else
        for(i = B; i >= 0; i--)
          ...
    }
\end{lstlisting}}
\end{minipage}
\hfill
\begin{minipage}[b]{0.43\linewidth}
  \centering
  \includegraphics[width=0.99\linewidth]{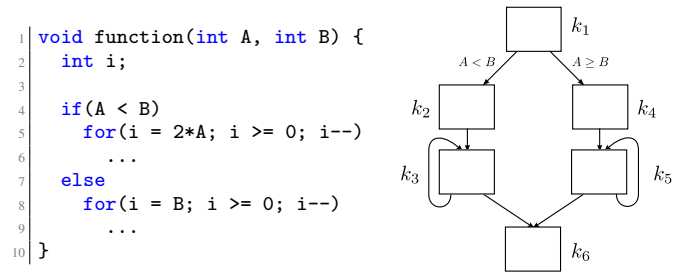}
  \label{fig:max_cfg}
\end{minipage}
\caption{\label{lst:max_example}Example program of where the analysis infers a \texttt{max} formula,
  together with its CFG}
\end{figure}

The results for the XMOS xCORE and ARM Cortex-M processors are shown in
Figures~\ref{fig:xmos_results} and~\ref{fig:arm_results}, respectively. These
graphs show the insertion sort, matrix multiplication and mac benchmarks, with
data series for the static analysis results and actual energy measurements. The static
analysis closely fits the empirical results, validating our
approach. Table~\ref{tab:table_results} shows the formulae and final errors for
all benchmarks. Overall, the final error is typically less than 10\% and 20\% on
the XMOS and ARM platforms respectively,
showing that the general trend of the static analysis results can be relied upon to give
an estimate of the energy consumption. We explain the sources of error in our results below:

\paragraph{Simple LLVM IR energy model (ARM).}
For the case of Cortex-M the errors in the analysis mostly stem from the greatly
simplified model of energy consumption in the Cortex-M. The LLVM energy model
used for the Cortex-M assigns an energy cost to each IR instruction. Therefore,
when an IR instruction expands to unexpected, or many ISA level instructions,
the energy consumption can be inaccurate. In particular, for base64, ternary
operators are heavily used inside its main loop. In LLVM IR, this introduces a
number of short conditional blocks inside this loop. These multiple basic blocks in LLVM
IR are translated to a smaller number of predicated instructions in the ARM ISA
by the compiler, so the static analysis will over approximate the energy consumed.

\paragraph{Measurement error.}
Measurement errors are introduced from environmental factors such as temperature
and power supply fluctuations. The tolerance of the components is also another
factor. The test harness contributes error too, as it must call a
function repeatedly in order to get its energy measurements. The loop
surrounding this function, together with the act of calling can be a significant
overhead when the amount of computation inside the function is low. In fact, we
can see that in all cases, the relative error converges to
a single error result. This is expected because in all of the benchmarks the
parameter controls the number of iterations performed in one or more loops. As
the parameter increases, the difference in the constant energy overhead is
minimized, with respect to the energy consumption of the function under test.
Measurement runs were run numerous times to ensure consistency of results within
the expected error margins described above.

\paragraph{Data flow through the processor's execution units.}
The energy models for the xCORE and ARM assume a random distribution of operand
data. In practice, however, operations such as logical tests, bit-manipulation
and instructions performed on shorter data types such as \texttt{char} will not
use the full bit-range of the data path. In cases such as these, energy consumption will
be lower, therefore introducing some estimation inaccuracy.

\paragraph{LLVM IR to ISA mapping (xCORE).}
In the case of the xCORE, the overall results are better than that of the
Cortex-M. This is due to a more accurate assignment of energy values to LLVM-IR
instructions, which the mapper can produce for each individual program, as
described in \secref{mapping}.
Nevertheless, the mapper introduces analysis error. For instance, the mapper does
not consider instruction scheduling on the processor, where an instruction fetch
stall can happen in some limited scenarios. This can be addressed by
performing a further local analysis on the ISA code to determine the possible
locations where this happens, and adjusting the energy accordingly.
Another problem arises when mapping LLVM IR \texttt{phi} instructions to the
``corresponding'' ISA code, in that ISA attributed to the \texttt{phi} may be
hoisted out of loops, but the \texttt{phi} is not. The hoisted ISA cost is
thus counted for each loop iteration, leading to an overestimate. This
phenomenon was partially addressed by automatically adjusting the energy of
\texttt{phi} instructions during the mapping.

\paragraph{Static analysis and data dependence.}
Programs where the behavior and state depends on complex properties of the
actual input data are problematic for static resource consumption
analysis. An extreme example of such a program would be an
interpreter. The execution time of an interpreter not only depends on the size
of the program file it is supplied, but also on the program represented in this
file. A more typical example would be the euclidean algorithm ($gcd(a,b)$),
where the number of 
steps taken to execute depends on a relationship between its parameters
$a$ and $b$. Our static analysis technique, however, still manages to compute an
approximation -- a logarithmic function with base $2$, which is dependent on
only one of the arguments. Part of the reason why we can analyze programs of
this type is that symbolic evaluation of modulus between two variables
$x \bmod y$ is defined as an upper bound of $y-1$, a lower bound of $0$ and an
approximation of $(y-1)/2$.

The \texttt{levenshtein} cost function for the xCORE processor includes a
\texttt{max} function, making it a different type of formula to the Cortex-M's
cost function. This occurs when a data dependent branch is on the upper bound of
the function and the analysis is unable to resolve the branch statically,
possibly because the branching is data dependent. An example of this is shown in
Figure~\ref{lst:max_example}. The analysis cannot statically ascertain the
outcome of the $A < B$ expression, so simply returns the cost function as the
maximum of the two possible branches:
\begin{equation*}
  function=k_1+\textrm{max}(k_2 + 2\cdot k_3\cdot A, k_4+k_5\cdot B)+k_6,
\end{equation*}
where $k_1,...,k_6$ are the costs of executing the respective basic blocks, as
seen in \figref{max_cfg}. The same effect causes
\texttt{max} to appear in the xCORE's formula --- there is a data dependent
\texttt{if} statement in an inner loop of \texttt{levenshtein}.

\subsection{Composability}
All of the benchmarks so far have consisted of relatively simple code, for which
a single function is analyzed. However, the analysis can handle nesting and
recursion, in the same way that it can handle functions with multiple basic
blocks. In the code in Listing~\ref{lst:composability_example}, the
\texttt{levenshtein} and modified \texttt{insertion sort} functions are composed into a
simple spell checker --- for a given string, sort
the list of strings by the \texttt{sortbysimilarity} to the target string.

\begin{lstlisting}[belowskip=-1pt, language=c,float,caption={Sort by similarity function,
    demonstrating that the analysis can be composed across multiple
    functions.},label=lst:composability_example]
int distances[MAX];

void sortbysimilarity(char *word, int word_len,
      char *dictionary[], int dictword_len,
      int n_strings)
{
    int i = n_strings;

    while(i--) {
      distances[i] = levenshtein(word,
              dictionary[i], word_len,
              dictword_len);
    }
    sort(distances, dictionary, n_strings);
}
\end{lstlisting}

In this listing, \texttt{dictword\_len} is the maximum size of the strings in
\texttt{dictionary}. Inferring a cost formula for this program does not present
any issues as long as it is possible to infer formulas for its constituent
parts. Our techniques construct Cost Relations (CRs) from the program that is being analyzed.
An important feature of CRs is their compositionality. This
allows computing closed form solutions of CRs composed of multiple relations by
concentrating on one relation at a time. The process starts by solving cost
relations that do not depend on any other relations and proceeds by replacing the
these cost relations in the equations which call such relations. For instance,
for the above program \texttt{levenshtein\_distance} has an associated energy
cost of
\begin{equation}
\left(A (53B+16)+35B+31\right)~\textrm{nJ},
\end{equation}
where $A$ and $B$ are the third and fourth arguments to the
function. Our modified string sorting routine has a cost of:
\begin{equation}
  \left(37 A^2+14 A+14\right)~\textrm{nJ}.
\end{equation}
These functions are systematically combined together so that a cost for
\texttt{sortbysimilary} is computed. In this case it is
\begin{equation}
\left(530 A B C+157 A C+346 BC+366 C^2+629 C+210\right)~\textrm{nJ},
\end{equation}
where $A$ is \texttt{word\_len}, $B$ is \texttt{dictword\_len} and $C$ is
\texttt{n\_strings}.

\section{Related Work}
Related work exists in four different areas: energy modeling of processors,
mapping low-level program segments to higher level structures, static resource consumption
usage analysis and worst-case execution time analysis (WCET).

Energy models of processors for program analysis require energy consumption data
in relation to the program's instructions. This data can be collected by
simulating the hardware at various levels, including
semiconductor~\cite{Nagel:M520} and CMOS~\cite{Bogliolo1997}.  Alternatively,
higher level representations may be used such as functional block
level~\cite{Steinke2001} that reflects the micro-architecture, direct
measurement on a per-instruction basis~\cite{Tiwari-embedded-1994}, or by
profiling the energy consumption of commonly used software
blocks~\cite{Qu-Function-level-2000}. Higher level data collection and modeling
efforts are typically quicker to use once the data has been acquired, as there
is less computational burden than a low-level simulation. However, the accuracy
may be lower, therefore a suitable trade-off must be met.

Although substantial effort has been devoted to ISA energy modeling, there is
not a lot of work done for higher level program representations. This is mostly
because precision decreases when moving further away from the hardware. One of
the most recently pertinent works for LLVM IR energy modeling is~\cite{Brandolese2011}.
The authors performed statistical analysis and characterization of LLVM IR code,
together with instrumentation and execution on the host machine, to estimate
the performance and energy requirements in embedded software. In their case,
retrieving the LLVM IR energy model to a new platform requires performing the
statistical analysis again. Our LLVM IR energy model takes into consideration
types and other aspects of the instructions. Furthermore, our mapping technique
requires only to adjust the LLVM mapping pass for the new architecture.

Static cost analysis techniques based on setting up and solving recurrence
equations date back to Wegbreit's \cite{DBLP:journals/cacm/Wegbreit75} seminal
paper, and have been developed significantly in subsequent
work~\cite{Rosendahl89, granularity, low-bounds-ilps97, vh-03, resource-iclp07,
AlbertAGP11a}. In~\cite{NMHLFM08} this approach is applied to inferring
statically the energy consumption of Java programs as functions of input data
sizes, by specializing a generic resource analyzer~\cite{resource-iclp07,
ciaopp-sas03-journal-scp} to Java bytecode analysis~\cite{resources-bytecode09}.
However, this work did not compare the results with measured energy
consumptions.  In~\cite{isa-energy-lopstr13} the approach is applied to the
energy analysis of XC programs using ISA-level models~\cite{Kerrison13}, and the
results are compared to actual hardware measurements.  Our analysis continues in
this line of work but with a number of important differences.  First, analysis
is performed at the LLVM-IR level and we propose novel techniques for reflecting
the ISA-level energy models at the LLVM-IR level.  Instead of using a
generic resource analyzer (requiring translating blocks to its Horn Clause-based
input syntax) and delegating the generation of cost equations to it, we generate
the equations directly from the LLVM-IR compiler representation, performing
control flow simplifications, and reducing the number of variables modelled by
the analysis mechanism.  Finally, we study a larger set of benchmarks.
Other approaches to cost analysis exist such as those using dependent
types~\cite{DBLP:journals/toplas/0002AH12}, SMT solvers~\cite{Alonso2012}, or
\emph{size change abstraction}~\cite{DBLP:journals/corr/abs-1203-5303}.

As discussed in~\secref{introduction}, energy and time are often correlated to
some degree. Techniques such as implicit path
enumeration~\cite{DBLP:conf/lctrts/LiM95} are often used in worst-case
execution time analysis of programs. In most cases, programs are assumed to be
preprocessed such that no loops are present (e.g. using loop unrolling). Some
approaches such as~\cite{DBLP:conf/vmcai/2009} focus on statically
predicting cache behavior. WCET analysis is concerned with getting an absolute
worst-case timing for hard real-time systems. In practice, for energy
consumption analysis we typically are more interested in average cases. Also,
most WCET analysis approaches produce absolute timing figures. In our case, we
infer energy formulae parameterized by the program's input.

\section{Conclusion and Future Work}
\label{sec:conc_future}
In this paper we have introduced an approach for estimating the energy
consumption of programs based on the LLVM compiler framework. We have shown
that this approach can be applied to multiple embedded languages (such as C or
XC), compiled using optimization level \texttt{O2} with different compilers
(such as Clang or XCC). We have also validated this approach for multiple
backends, via two target architectures: ARM Cortex M3 and XMOS XS1-L. Our
approach is validated by comparing the static analysis to physical measurement
taken from the hardware. The results on our benchmarks show that energy
estimations using our technique are within 10\% and 20\% or better in the case
of the xCORE and the Cortex-M processors, respectively.

Although the techniques discussed here were initially designed for single
threaded programs, these can be adapted to multi-threaded programs. To do so,
we need to take the synchronization time into consideration. For
example, the XC language has explicit constructs for thread communication using
channels, and
therefore the blocking communication between threads needs to be modeled. In
order to do so, we can analyze the communication throughput of individual
threads using techniques discussed in this paper. Using this information we can
estimate the time between events happening on channels and hence the utilization
of the processor. This, coupled with multi-threaded energy models as discussed
in~\secref{xs1model}, can be used to analyze multi-threaded programs.

An interesting direction is to further develop the assignment of
energy to LLVM IR program segments. In particular, an LLVM IR energy model for
the xCORE can be implemented by using the information gathered from the mapping
technique together with statistical analysis. The mapping technique used for the xCORE
can also be adapted for the ARM case. We  aim to further develop our
techniques so they can be applied against other embedded
processor architectures, such as MIPS, or other ARM variants.

Finally, the static analysis techniques can be improved further. Currently the biggest
limitation is solving the cost relations. Cost relations could also be
solved numerically, enabling us to analyze more complex programs. An
implementation of this can be used when actual formulae are not
required.

\acks
The research leading to these results has received funding from the European
Union 7th Framework Programme (FP7/2007-2013) under grant agreement no 318337,
ENTRA - Whole-Systems Energy Transparency. Special thanks are due to Pedro
Lopez-Garcia and his team at the IMDEA Software Institute for many fruitful
discussions and inspiration. We would like to thank our project
partners at Roskilde University and at XMOS. Thanks also go to Samir Genaim
for his help on how to best make use of the PUBS solver.

\bibliographystyle{abbrv}
\bibliography{entra}
\end{document}